\documentclass{appolb}
\usepackage{graphicx}
\input epsf

\begin{document}
\title{Eta-mesic nucleus and COSY-GEM data
\thanks{Presented at the International
Symposium on Mesic Nuclei, Krak\'ow, Poland, June 16, 2010.}}
\author{ Q. Haider
\address{Physics Department, Fordham University\\
Bronx, N.Y. 10458, USA}
\and
Lon-Chang Liu
\address{Theoretical Division, Group T-2, Los Alamos National Laboratory\\
Los Alamos, N.M 87545, USA}}
\date{\today}
\maketitle

\vspace{1cm}
\begin{abstract}

The experimental data of the COSY-GEM collaboration for the recoil-free transfer reaction $p(^{27}$Al,$^{3}$He)$\pi^{-} p' X$,
leading to the formation of  bound state of eta $(\eta$) meson in $^{25}$Mg nucleus, is reanalyzed in this paper. In particular,
predicted values of binding energy and half-width of the $\eta$-mesic nucleus $^{25}$Mg$_{\eta}$, given by different theoretical
approaches, are compared with the ones obtained from the experimental missing mass spectrum. It is found that the spectrum can be
explained reasonably well if interference effect of another process, where $\eta$ is not bound in $^{25}$Mg but is scattered by
the nucleus and emerge as a pion, is taken into account. The data also indicate that the interaction between N$^{*}$(1535) and
a nucleus is attractive in nature.

\end{abstract}


\section{Introduction}{\label {sec.1} }
The existence of eta-mesic nucleus, a bound state of $\eta$ meson in a nuclear orbital, was first predicted by us in 1986~\cite{hai1}.
The formation of such a nucleus is a consequence of the attractive $\eta$-nucleon ($\eta$N) interaction in the threshold region~\cite{bhal}.
This attractive nature of the $\eta$N interaction is due to the fact that the N$^{*}$(1535) resonance is situated not too far above
the threshold of the $\eta$N channel ($\sim$ 1488 MeV). Our prediction has been corroborated by other researchers and can be
found in the vast amount of theoretical work published during the past 20 years~\cite{naga, wilk}.

Experimental efforts to detect $\eta$-mesic nucleus, on the other hand, did not yield positive results~\cite{chri, john}. It was only recently
the COSY-GEM collaboration~\cite{budz} confirmed unambiguously the detection of mesic-nucleus in the recoil-free transfer reaction
$p(^{27}$Al,$^{3}$He$)\pi^{-}p'X$. The experimental kinematics was chosen such that in the first stage of the
reaction the $\eta$ was produced nearly at rest, thus favoring its capture by the residual nucleus $^{25}$Mg to form
the $\eta$-mesic nucleus that we denote as $^{25}$Mg$_{\eta}$: \\

\hspace{0.75in} $p \;+\; ^{27}$Al $\rightarrow (\eta \;+\; ^{25}$Mg$) \;+\; ^{3}$He $\rightarrow \; ^{25}$Mg$_{\eta} \;+\; ^{3}$He.\\

\noindent
Because of energy conservation, the bound $\eta$ cannot reappear
as an observable in the decay products. Rather it interacts with a target nucleon resulting in the emission of a nearly ``back-to-back"
$\pi^{-}p$ pair in the laboratory, {\it i.e.},

\hspace{0.75in} $^{25}$Mg$_{\eta} \rightarrow (\eta \;+\; n) \;+\; ^{24}$Mg $\rightarrow (\pi^{-} \;+\; p) \;+\; X.$\\

\noindent We may call the above multi-step reaction Process M as it proceeds through the doorway state containing mesic nucleus. By fitting
the missing-mass spectrum with $|f_{b}|^{2} + |f_{g}|^{2}$, where $f_{b}$ and $f_{g}$  are, respectively, the background amplitude
and a Gaussian function, the COSY-GEM collaboration has determined the values of the binding energy and FWHM of the mesic nucleus $^{25}$Mg$_{\eta}$
to be, respectively, $\epsilon = (- 13.13 \pm 1.64)$~MeV and $\Gamma = (10.22 \pm 2.98)$~MeV (or $\Gamma/2\simeq 5.1 \pm 1.5$~MeV.
We refer the readers to ref.~\cite{budz} for the experimental details and the accompanying analysis of the data.

The experimental verification of the existence of mesic-nucleus by COSY-GEM collaboration has opened up new avenues in the study of exotic nuclei.
In this work, we reanalyze the data and investigate how theory measures up to the experimental results. In particular,
we want to address the following questions:

\begin{enumerate}
\item Are the published values of s-wave $\eta$N scattering length ($a_{\eta N}$) consistent with the data?
\item Can the data be described within the context of the models and formalism used by us~\cite{hai2, hai3}, which  predicted
the existence of $\eta$-mesic nucleus?
\item Is it possible that under the conditions the experiment was done, $\eta$ got captured in the excited state of $^{25}$Mg?
\item Is there any other process that could have contributed to the observed experimental spectrum?
\item Can the observed spectrum provide information about the nature
of the N$^{*}$(1535)-nucleus interaction?
\end{enumerate}
We believe that answer to the above questions will enhance our understanding of $\eta$-nucleus reaction dynamics
and properties
of $\eta$-mesic nucleus.

\section{Theoretical analysis}{\label{sec.2}}
To obtain the binding energy and half-width of $\eta$-mesic nucleus, we solve the relativistic three-dimensional (covariant) integral equation
\begin{equation}
\frac{{\bf k'}^{2}}{2\mu}\;{\psi}({\bf k'})
+ \int\;d{\bf k}<{\bf k}'\mid {V}\mid{\bf k}>{\psi}({\bf k})
=E{\psi}({\bf k}')\ ,
\label{eq:1}
\end{equation}
where ${\bf k}$ and ${\bf k}'$ are the momenta of $\eta$ in the $\eta$-nucleus c.m. frame, and $\mu$ is
the reduced mass of the $\eta$-nucleus system. The eigenvalue $E$ is complex and is given by
$E=\epsilon-i\Gamma/2$ with $\epsilon < 0$ and $\Gamma > 0$.

\subsection{Scattering-length approach}{\label{sec.2.1}}

In the scattering-length approach, the first-order low-energy $\eta$-nucleus optical potential is
\begin{equation}
<{\bf k}'|V|{\bf k}> = -\frac{1}{4\pi^{2}\mu}\left ( 1+\frac{M_{\eta}}{M_{N}}\right )a_{\eta N}f ({\bf k}-{\bf k}'),
\label{eq:2}
\end{equation}
where $f({\bf k} - {\bf k}�)$ is the nuclear form factor, $M_{\eta}$ is the mass of $\eta$ meson, and $M_{N}$ is the nucleon mass.
The only input in calculating  $\epsilon$ and $\Gamma/2$ is, therefore, the scattering length $a_{\eta N}$. This approach can be also termed
as ``on-shell" calculation because the scattering length is defined at $\eta$N on-shell threshold energy $M_{\eta}+M_{N}$.

The value of $a_{\eta N}$ that reproduces the $\epsilon$ and $\Gamma/2$ determined by the COSY-GEM analysis equals to $(0.292 + 0.077i)$~fm.
We have noted, however, that all published theoretical models give $a_{\eta N}$ having an imaginary part that is at least a factor of 2.4
greater than 0.077 fm. The existence of this significant difference is theoretical-model independent and
needs to be understood. In the next subsection we analyze the data more thoroughly by using a microscopic optical
potential which takes into account the energy dependence of $\eta$N interaction below the $\eta$N threshold.

\subsection{Microscopic (off-shell) optical potential approach}{\label{sec:2.2}}

The momentum-space matrix elements of the microscopic $\eta$-nucleus optical potential is~\cite{hai1, hai2}
\begin{equation}
<{\bf k}'|V|{\bf k}> = <{\bf \kappa}'|t_{\eta N}(W)|{\bf \kappa }>.
\label{eq:3}
\end{equation}
The variables $\kappa$ and $\kappa '$ are the initial and final $\eta$N relative momenta. The total energy of the $\eta$N system in its
center-of-mass frame is
\begin{equation}
W = M_{\eta} + M_{N} + <B_{N}>,
\label{eq:4}
\end{equation}
where $<B_{N}> \;(< 0)$ is the average binding energy of the nucleon. We have calculated the matrix elements of the scattering operator
$t_{\eta N}(W)$ by using the coupled-channel isobar (CCI) model of Bhalerao and Liu~\cite{bhal}, which has its parameters
determined from fitting the $\pi$N S$_{11}$ phase shifts. The same CCI model was used in all of our earlier calculations that predicted
the existence of $\eta$-mesic nucleus. More importantly, the model  provides us with the detailed energy dependence of the interaction
in the energy region where $\eta$ can be bound in a nucleus. It contains strong-interaction form factors and satisfies off-shell unitarity.

Because for bound-state problems the contributions of the $p$- and $d$-wave $\eta$N interactions are negligibly small,
only the $s$-wave $\eta$N interaction needs to be taken into account. The $t$-matrix of the CCI model, therefore, has the form
\begin{equation}
<{\bf \kappa}'|t_{\eta N}(W)|{\bf \kappa }>=Kv({\bf \kappa}',\Lambda ){\cal A}(W)v({\bf \kappa },\Lambda ),
\label{eq:5}
\end{equation}
where $K$ is a kinematic factor, $v$ is the $s$-wave off-shell form factor, and $\Lambda$  is the range parameter.
The energy-dependent amplitude is
\begin{equation}
{\cal A}(W) = \frac{g^{2}}{2W D(W)},
\label{eq:6}
\end{equation}
where $g$ is the $\eta$NN$^{*}$ coupling constant. The expression for $D(W)$ is
\begin{equation}
 D(W) = W - \{ M^{0} + V_{N^{*}}(W) + \Sigma^{free}(W) + i {\cal I}m[\Sigma^{abs}(W)] \},
 \label{eq:7}
\end{equation}
where $M^{0}$  is the bare mass of the resonance, and the N$^*$(1535)-nucleus interaction is defined as
\begin{equation}
V_{N^{*}}(W) = r(W) + {\cal R}e[\Sigma^{abs}(W)].
\label{eq:8}
\end{equation}
In the above expressions, $\Sigma^{free}(W)$ is the complex self-energy of N$^{*}$ arising from its decay to the $\eta$N, $\pi$N, and
$\pi\pi$N channels. The self-energy arising from absorption or annihilation of the $\eta$ and $\pi$ resulting
from N$^{*}$ decaying into $\eta$N, $\pi$N, and $\pi\pi$N is denoted by $\Sigma^{abs}(W)$. The $r(W)$ in eq.(\ref{eq:8}) is a real quantity
which accounts for all other subthreshold N$^*$-nucleus interactions. Although ${\cal I}m[\Sigma^{abs}(W)]$ has been evaluated within the
framework of local density approximation~\cite{chia}, theoretical models for $r(W)$ as well as for calculating ${\cal R}e[\Sigma^{abs}(W)]$
that can be systematically checked against data are still unavailable. We, therefore, treat $V_{N^*}(W)$ as a parameter.

We emphasize that unitarity of an optical potential requires that the imaginary parts of $\Sigma^{abs}(W)$ and $\Sigma^{free}(W)$ should have
the same sign. Consequently, $\mid {\cal I}m[D(W)]\mid \geq \mid {\cal I}m[\Sigma^{free}(W)]\mid$. In our calculations, careful attention
is paid to this requirement.

Upon applying the CCI model to $^{25}$Mg with $V_{N^{*}}(W)$=0 and $<B_N>=-30$~MeV, we obtained binding energy $\epsilon =-6.5$~MeV
and half-width $\Gamma/2 = 7.1~$MeV. (We point out that $<B_N> = -30$~MeV is based on the findings obtained from extensive pion-nucleus
studies in the literature.) As we can see, although the calculated half-width is comparable with the data, the
magnitude of the calculated binding energy is about 7~MeV too small.

To reproduce the COSY-GEM values of $\epsilon=- 13$~MeV and $\Gamma/2 = 5.0$~MeV requires that the amplitude ${\cal A}=-(0.0521+0.0099i)$~fm$^{2}$.
A knowledge of the value of ${\cal A}$
allows us to determine $W$, and hence $<B_{N}>$, by means of two equations for ${\cal I}m[D(W)]$ obtained from eqs.({\ref{eq:6}) and
(\ref{eq:7}). They are
\begin{equation}
 {\cal I}m[D(W)] = \left (\frac{g^{2}}{2W} \right ){\cal I}m\left(\frac{1}{{\cal A}(W)}\right),
\label{eq:9}
\end{equation}
and
\begin{equation}
 {\cal I}m[D(W)] = -\{ {\cal I}m[\Sigma^{free}(W)] + {\cal I}m[\Sigma^{abs}(W)]\}.
\label{eq:10}
\end{equation}
With the fitted value of ${\cal A}$ and the calculated imaginary part of the self energies, we find that the values of ${\cal I}m[D(W)]$ given
by the last two equations are same if $W=1125$~MeV. This implies $<B_{N}>=-360$~MeV, which is clearly unrealistic. We conclude, therefore, that
a simple application of optical potential does not yield satisfactory result, and a reexamination of the whole reaction is in order.

\subsection{Excited state of $^{25}$Mg}{\label{sec.2.3}}

In view of the above findings, it is reasonable to investigate whether the seemingly large binding energy of $^{25}$Mg$_{\eta}$ could
be due to the binding of $\eta$ on excited states of $^{25}$Mg. To answer this question, we recall that the experimental spectrum
as a function of the binding energy $E$ of $\eta$ is deduced from measurements of the missing mass $\Delta M$. In terms
of the mass $M_{25}$ and excitation energy $E_{x} \;(> 0)$ of $^{25}$Mg, energy conservation and recoil-free kinematics give
\begin{equation}
\Delta M = M_{25} + M_{\eta} + E_{x} + E.
\label{eq:9}
\end{equation}
With known values of the masses, we find that the experimental centroid of the spectrum at $\Delta M = 23.803$~GeV~\cite{budz} leads to
\begin{equation}
E = - (13 + E_{x}) \;\; \mbox{(MeV)}.
\label{eq:10}
\end{equation}
From this equation, we see that the centroid of the observed spectrum at $-13$~MeV implies $E_{x} = 0$.
We, therefore, conclude that the $\eta$ was captured in the ground-state of $^{25}$Mg and that the noted
difference between theory and experimental data cannot be due to $\eta$ being bound on excited states.

\section{Two doorway processes}{\label{sec.4}}

The inability of the theoretical models to explain the data within the framework of doorway process M alone leads us to investigate whether there
is another process contributing to the observed spectrum but has been overlooked in the above discussions. Indeed, the $\eta$ produced in the
intermediate state can scatter from the residual nucleus and emerge as a pion, without going through formation of an $\eta$-mesic nucleus.
This latter multi-step process is

\hspace{0.75in} $p \;+\; ^{27}$Al $\rightarrow (\eta \;+\; ^{25}$Mg$) \;+\; ^{3}$He\\

\noindent
followed by

\hspace{0.75in} $\eta \;+\; ^{25}$Mg $\rightarrow (\eta \;+\; n) \;+\; ^{24}$Mg $\rightarrow  (\pi^{-} \;+\; p) \;+\; X.$\\

\noindent
We call this Process S (S for scattering). It should be emphasized that because the Processes M and S lead to the same final experimental state,
the effect of interference between them must be taken into account. We, therefore, fit the experimental spectrum with
the function $\alpha |f_{S} + f_{M}|^{2}$, where $f_{S}$  and $f_{M}$ are the amplitudes for the Processes S and M, respectively, and $\alpha$
adjusts the overall magnitude of the spectrum. The amplitudes are parameterized as
\begin{equation}
f_{S} = \lambda e^{i\theta}, \;\;\;\; f_{M} = - \; \frac{\Gamma/2}{E-(\epsilon -i\Gamma/2)}.
\label{eq:10}
\end{equation}
The parameters $\lambda$ and $\theta$ represent the relative strength and phase between the two processes at $E=\epsilon$. The energy dependence
of $f_{S}$ is neglected in the present phenomenological approach because it is much smoother than $f_{M}$. The half-width controls the sharpness
of the peak's structure of the calculated spectrum. With the aid of eq.(\ref{eq:10}), we carried out fits to the data by varying $\alpha , \; \lambda$,
and $\theta$ while keeping fixed the values of $\epsilon$ and $\Gamma/2$, which were calculated using the microscopic optical potential.
The fits for three different sets of parameters are shown in figure~1. The values of the parameters are listed in table~\ref{tab1}.

\begin{figure}
{\centerline{\includegraphics[angle=0,width=0.5\columnwidth]{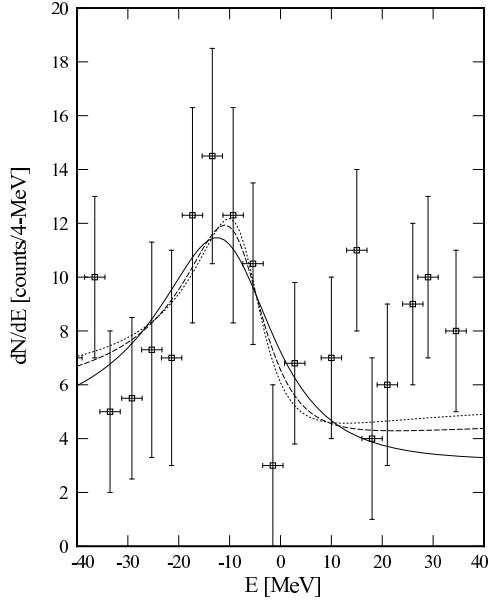}}}
\caption{Spectra obtained with $\epsilon-i\Gamma/2$ fixed at :
(a) $-(6.5 + 7.1i)$~MeV (dotted curve); (b) $-(8.0 + 9.6i)$~MeV (dashed curve);
(c)~$-(10.0 + 14.0i)$~MeV (solid curve). The data
are from ref.~\cite{budz}.}
\label{Fig2}
\end{figure}

\begin{table}
\begin{center}
\caption{ Fitted values of the parameters.}
\label{tab1}

\medskip
\begin{tabular}
 {|c|c|c|c|c|c|c|c|c|}\hline\hline
Fit & $\epsilon$       & $\Gamma /2$       & $\alpha$ & $\lambda$ & $\theta$ & $W$& $<B_{N}>$ & $V_{N^{*}}$ \\ \hline
(a) &  $-6.5$                 & $7.1$                   & $1.83$   & $1.75$ & $0.62$   &   $1458$ & $-30$ & $0$\\
(b) & $-8.0 $                 & $9.6$                   & $1.91$   & $1.14$ & $0.79$   &   $1458$ & $-30$ & $-15$ \\
(c) &$-10.0$                  & $14.1$                  & $2.54$   & $1.17$ & $1.06$   &   $1458$ & $-30$ & $-42$ \\ \hline
\end{tabular}
\end{center}
($\epsilon , \; \Gamma /2, \; W, \; <B_{N}>$, and $V_{N^{*}}$ are in MeV,
$\alpha$ is in counts/4-MeV, $\lambda$ is dimensionless, and $\theta$ is in radians.)
\end{table}

A few remarks about the fits are in order. First, fit (a) is obtained without nuclear medium effect
({\it i.e.,} $V_{N^{*}} = \Sigma^{abs}(W)=0$, as in our previous predictive calculations), but with the interference effect between
$f_{S}$ and $f_{M}$ included. As can be seen, interference shifted the observed peak from $-6.5$~MeV to $-9.0$~MeV. Fits (b) and (c),
on the other hand, are obtained with $V_{N^{*}}<0.$ One can see that the final positions of the peak of $\alpha |f_{S} + f_{M}|^{2}$ are
approximately at $-11$~MeV for fit (b) and $-13$~MeV for fit (c). In general, the interference effect causes the final peak positions,
as indicated by the respective curves in figure~1, to move closer to the experimental centroid. In fact, in the case of fit (c) it is equal
to the experimental centroid. The shift in the position of the peaks clearly indicates the importance of the effects of interference.
The negative value of $V_{N^{*}}$ for fits (b) and (c) indicates that the data favor an attractive N$^{*}$(1535)-nucleus interaction.

One may ask why we did not fit the data by using larger $|\epsilon |$ and smaller $\Gamma/2$ or stronger $V_{N^{*}}$?
Without going into details, we just mention that these quantities cannot be treated as completely free parameters.
As pointed out in section \ref{sec:2.2}, their relative strengths are constrained by the unitarity requirement of the optical-potential:
$\mid{\cal I}m[D(W)]\mid \geq \mid {\cal I}m[\Sigma^{free}(W)]\mid$.

\section{Conclusion}{\label{sec:3}}

Our analysis shows that two reaction processes are contributing to the observed spectrum of bound $\eta$ in $^{25}$Mg. The quantum interference
between the two processes results in a binding of $\eta$ that is weaker than suggested by the centroid of the observed spectrum. The present
analysis gives the binding energy between $-8$ and $-10$~MeV and the half-width between 10 and 14~MeV if $^{25}$Mg is in its ground state.

The present analysis also indicates that the real part of interaction between N$^{*}$(1535) and a medium-mass nucleus is attractive at energies
below the $\eta$N threshold. This
latter information should be of value to nuclear physics studies involving the baryon resonance~N$^{*}$(1535).

Microscopic calculation of $f_S$ and $f_M$ is in progress, where we calculate both amplitudes by using the same
$\eta$-nucleus optical potential. Accordingly, there is no need of the phenomenological parameters $\lambda$ and $\theta$.
Preliminary results reaffirm the basic findings reported here, namely, (i) the existence of two reaction processes,
(ii) the interference effect between them causes the peak of $\mid f_S + f_M\mid^2$ to appear at a binding energy
much stronger than the actual binding energy of $^{25}$Mg$_{\eta}$, and (iii) the attractive nature of the N$^{*}$-nucleus interaction.
Details will be reported in a future publication.

\medskip
One of us (Q.H.) would like to thank Dr. Pawel Moskal for the hospitality extended while he was in Krak\'ow, Poland to attend
the International Symposium on Mesic Nuclei.\\

\pagebreak


\begin{thebibliography}{5}
\bibitem{hai1}	Q. Haider and L.C. Liu, {\it Phys. Lett.}  {\bf B172}, 257 (1986); {\bf B174}, 465E (1986).
\bibitem{bhal}	R.S. Bhalerao and L.C. Liu, {\it Phys. Rev. Lett.} {\bf 54}, 865 (1985).
\bibitem{naga}	H. Nagahiro, M. Takizawa, and S. Hirenzaki, {\it Phys. Rev. C} {\bf 74}, 045203 (2006) and references therein.
\bibitem{wilk}	C. Wilkin {\it et al.}, {\it Phys. Lett.} {\bf B654}, 92 (2007) and references therein.
\bibitem{chri}  R.E. Chrien {\it et al.}, {\it Phys. Rev. Lett.} {\bf 60}, 2595 (1988).
\bibitem{john}	J.D. Johnson {\it et al.}, {\it Phys. Rev.  C} {\bf 47}, 2571 (1993).
\bibitem{budz}	A. Budzanowski {\it et al.}, {\it Phys. Rev. C} {\bf 79}, 012201(R) (2009).
\bibitem{hai2}	Q. Haider and L.C. Liu, {\it Phys. Rev. C} {\bf 66}, 045208 (2002).
\bibitem{hai3}	Quamrul Haider and Lon-chang Liu, {\it Acta Phys. Pol. (Proceedings Supplement)} {\bf B2}, 121 (2009).
\bibitem{chia}  H.C. Chiang, E. Oset, and L.C. Liu, {\it Phys. Rev. C} {\bf 44}, 738 (1991).
\end{thebibliography}
\end{document}